# Design and integration of a crab-waist interaction region for the Super Tau-Charm Facility

Linhao Zhang[1,*,ξ], Tao Liu[1,*], Ye Zou[1,ξ], Penghui Yang[2], Demin Zhou[3], Jiancong Bao[1], Ze Yu[1], Yuhan Jin[1], Yihao Mo[1], Sangya Li[2], Tianlong He[2], Qing Luo[1,2], Jingyu Tang[1,2]

[1] School of Nuclear Science and Technology, University of Science and Technology of China, No. 443 Huangshan Road, Hefei, 230027, Anhui, China
[2] National Synchrotron Radiation Laboratory, University of Science and Technology of China, No. 42, South Cooperative Road, Hefei, 230029, Anhui, China
[3] KEK, Oho 1-1, Tsukuba, 305-0801, Ibaraki, Japan

The Super Tau-Charm Facility (STCF) is a new-generation e+/e− collider proposed in China, designed to operate in the center-of-mass energy range of 2-7 GeV. To achieve the design luminosity exceeding 5×10$^{34}$ cm$^{−2}$s$^{−1}$ at the optimal energy of 4 GeV, a large crossing angle combined with the crab-waist correction scheme is adopted. However, this collision scheme introduces strong nonlinearities in the interaction region (IR) due to the extremely low vertical beta function of $\beta_y^* \leq 1$ mm, which significantly limits dynamic and momentum apertures of the collider ring, especially when nonlinear effects interfere with the crab sextupoles. This paper presents a comprehensive modular optics design that addresses these challenges through several key features: (1) local chromaticity correction up to third order to enhance momentum bandwidth; (2) exact -I transformation between chromatic sextupole pairs for nonlinear cancellation; (3) minimization of the dispersion invariant along the IR to improve local momentum acceptance; and (4) optimized beta functions at crab sextupole locations to reduce strength requirements and associated nonlinearities. Resonance driving terms analysis confirms effective suppression of geometric aberrations while preserving the intended crab-waist effects. When integrated into the collider ring, the design achieves a Touschek lifetime exceeding 300 s at beam energy of 2 GeV, satisfying STCF requirements. The impact of fringe fields from superconducting quadrupoles is mitigated using octupole correctors, and detector solenoid effects are fully suppressed via local anti-solenoid compensation. Furthermore, the defined machine-detector interface layout ensures minimal synchrotron radiation background at the central beryllium chamber, achieved through optimized beam-pipe transition and outer-ring injection geometry, while ultra-high vacuum conditions are required to suppress beam-gas background. This IR design represents the current optimized solution for STCF and has been incorporated into the project's conceptual design report.

## 1. Introduction

---

[*] These authors (Linhao Zhang and Tao Liu) contributed equally to this work.
[ξ] Corresponding authors: zhanglinhao@ustc.edu.cn, zouye@ustc.edu.cn

Super Tau Charm Facility (STCF), a new-generation electron-positron collider currently under development in China, is designed to operate at center-of-mass (CoM) energies ranging from 2 to 7 GeV [1]. Its design luminosity targets at least $5\times10^{34}$ cm$^{-2}$s$^{-1}$ at the optimal CoM energy of 4 GeV, which is about 50 times that of the currently operating Tau-Charm factory—BEPCII. The STCF will serve as a unique platform for precision physics studies in the tau-charm energy region and d potentially explore new physics beyond the standard model.

To achieve the desired high luminosity, STCF adopts the novel collision scheme proposed by P. Raimondi in 2006, which combines a large Piwinski angle with crab-waist correction [2]. This scheme has been successfully validated at the DAΦNE Φ factory [3]. With the large Piwinski angle $\phi_{\text{Piw}}$, defined by $\phi_{\text{Piw}} = \sigma_z/\sigma_x \cdot \tan(\theta_c)$ where $\theta_c$ is half of the full crossing angle, $\sigma_z$ and $\sigma_x$ are the RMS bunch length and horizontal size, the overlap area of colliding bunches is much smaller than the bunch length—allowing the vertical beta function at the interaction point (IP), $\beta_y^*$, to be significantly squeezed into $\sigma_z/\phi_{\text{Piw}}$ without significant hourglass effects. However, the large Piwinski angle also excites strong adverse betatron and synchrobetatron beam-beam resonances. These resonances can be effectively suppressed by a pair of crab sextupoles [4], placed with specific phase advances to IP ($\Delta\mu_x = m\pi, \Delta\mu_y = (2n+1)\pi/2$) and tailored strength, which shift the vertical waist of one beam onto the central axis of the opposing beam at the IP.

Globally, this collision scheme has been adopted by all new-generation e+/e- colliders to achieve "super-factory" luminosities for precise particle physics study in their respective energy regions. For example, in the high-energy region, the under-study Circular Electron Positron Collider (CEPC) [5] in China and Future Circular Collider (FCC-ee) [6] in Europe focus on precise Higgs boson studies (acting as Higgs factories) while also functioning as Z, W, and top-quark factories. In the medium-energy range, SuperB was previously proposed by INFN and SLAC [7] but terminated now, while SuperKEKB at KEK in Japan—commissioned since 2016—recently achieved a record luminosity of $5.1\times10^{34}$ cm$^{-2}$s$^{-1}$ in 2025 [8]. In the low-energy tau-charm region, the ongoing super tau-charm facility plans include SCTF from BINP in Russia [9] and STCF at USTC in China [10], following the early proposed Tau/Charm Factory at INFN in Italy (discontinued now) [11].

However, the crab-waist collision scheme poses substantial challenges to the IR optics design [12]. Key issues include: (1) Extremely low $\beta_y^*$ leads to excessively large beta-functions at the final focus (FF) quadrupole doublets, requiring ultra-strong FF quadrupoles. These quadrupoles generate large natural chromaticity (needing strong sextupoles for local chromaticity correction), induce strong fringe fields, and increase sensitivity to field errors; (2) Extremely low $\beta_y^*$ also introduces unavoidable high-order kinematic terms of IP drift; (3) Crab sextupoles demand specific phase advances from the IP. Any lattice nonlinearity or imperfection between the sextupoles disrupts these phase constraints; (4) Compensating the detector solenoid field with anti-solenoids is essential, as the solenoid introduces horizontal and vertical coupling. All these IR-related nonlinear and high-order effects significantly degrade the collider ring's dynamic and momentum apertures, causing a very short Touschek lifetime for the STCF collider ring under high bunch currents and low emittances—for example, less than 300 s, much shorter than that of other circular electron accelerators. Operational and commissioning experience from SuperKEKB further confirms that most unavoidable lattice nonlinearities in collider rings originate from the IR [13-15]. Additionally, achieving extremely small $\beta_y^*$ requires ultra-strong superconducting FF quadrupoles within limited space, imposing strict constraints on machine-detector interface (MDI) layout design. Therefore, coordinating these combined effects to meet the collider's design luminosity constitutes the central challenge in crab-waist IR design.

An initial preliminary IR lattice design for STCF was presented in 2019 [16]. After achieving first-order and second-order local chromaticity correction, the IR was integrated into the STCF ring (circumference ~ 575 m), but yielded extremely small dynamic and momentum apertures, leading to a short Touschek lifetime of 35 s [17]. This STCF ring optics was later refined to include double rings, damping wigglers, and injection sections [18], but the IR layout and optics saw no significant improvements beyond further correction of high-order chromaticity within IR. Following careful global nonlinear optimization, this updated ring (circumference ~ 617 m) achieved larger dynamic and momentum apertures, with a Touschek lifetime of 240 s at a luminosity of $5.1 \times 10^{34}$ cm$^{-2}$s$^{-1}$ and the beam energy of 2 GeV [19]. Notably, both early lattice versions prioritized longitudinal beam polarization via Siberian snakes—imposing strict layout constraints due to the required specific bending angles between successive snakes. Since January 2024, polarization has been deprioritized, enabling significant redesign of the collider ring layout along with the arc and IR optics, while retaining the previous lattice as a backup [1]. The IR layout and optics have been specifically optimized to enhance nonlinear dynamics performance, which will be described in this paper. Upon integration into the latest collider ring, the design achieves a Touschek lifetime exceeding 300 s at a luminosity of $9.4 \times 10^{34}$ cm$^{-2}$s$^{-1}$ and a beam energy of 2 GeV. Further details on the latest collider ring optics are available in Ref. [20].

This paper focuses on the physics design of the STCF crab-waist IR, covering the selection of IR key parameters, modular IR optics design and local chromaticity correction, IR performance in the collider ring, and the MDI layout and physics-related design (including synchrotron radiation, beam-stay clear and detector background). The proposed IR design herein represents the current optimal solution for the STCF collider ring, and has been incorporated into the STCF conceptual design report [21].

**2. IR key parameter selection: a physics-driven trade-off**

The design of the IR is governed by several key parameters that directly influence the luminosity, nonlinear dynamics, and engineering feasibility. The selection of these parameters is a critical step that involves balancing competing physical effects and constraints. We consider the full crossing angle ($2\theta_c$), the drift length from the IP to the first FF quadrupole ($L^*$), and the horizontal and vertical beta functions at the IP ($\beta_x^*$ and $\beta_y^*$). Table 1 compares these parameters with those of other new-generation e+/e- colliders.

- Crossing angle $2\theta_c$:

    A large crossing angle (e.g., 60 mrad) is beneficial for achieving a large Piwinski angle, which allows for a reduced vertical beta function without the hourglass effect. It also enables rapid beam separation and reduces the strength of the crab sextupoles, thereby mitigating their nonlinear impact. Furthermore, a large crossing angle lowers the horizontal beam-beam parameter ($\xi_x$), which helps satisfy the condition $v_z \gg \xi_x$ ($v_z$ is the synchrotron tune) for suppressing the coherent X-Z beam-beam instability [22, 23]. However, a larger crossing angle leads to geometric luminosity loss owing to reduced vertical beam-beam parameter ($\xi_y$) and necessitates higher beam currents to meet the luminosity target, thereby posing greater challenges from collective effects. Therefore, the choice of $2\theta_c$ represents a trade-off between luminosity and beam-beam stability. We have selected $2\theta_c = 60$ mrad for STCF, while the feasibility of a larger angle (e.g., 80 mrad) remains under investigation.

- $L^*$:

    The drift length $L^*$ from the IP to the first FF quadrupole affects both the chromaticity and the MDI space. A shorter $L^*$ reduces the natural chromaticity of the IR, easing the requirement for

chromaticity correction. However, it also constrains the space available for the detector and the dual-aperture FF superconducting quadrupoles, complicating their design and installation. With a crossing angle of 60 mrad, we have chosen $L^* = 0.9$ m to provide sufficient space for the quadrupoles while maintaining acceptable chromaticity.

- $\beta_y^*$:

The vertical beta function at the IP is a primary factor for achieving high luminosity. A smaller $\beta_y^*$ directly increases the luminosity but requires stronger FF quadrupoles. These stronger quadrupoles, in turn, introduce larger natural chromaticity and fringe field nonlinearities. Therefore, the choice of $\beta_y^*$ involves a trade-off between luminosity and nonlinear effects. To reach the STCF design luminosity of $1\times10^{35}$ cm$^{-2}$s$^{-1}$ (twice the engineering goal of $5\times10^{34}$ cm$^{-2}$s$^{-1}$), $\beta_y^*$ must be below 1 mm, considering the vertical beam-beam parameter below 0.1 and beam-intensity effects. Our current design adopts $\beta_y^* = 0.8$ mm, instead of the previously considered 0.6 mm [19], aligning with lowest operational level achieved to date by SuperKEKB [24]. This choice helps alleviate IR chromaticity and nonlinearity while maintaining high luminosity.

- $\beta_x^*$:

The horizontal beta function at the IP influences the horizontal beam-beam parameter and the IR chromaticity. A smaller $\beta_x^*$ helps reduce $\xi_x$, which is beneficial for suppressing the coherent X-Z instability. However, it also increases the horizontal chromaticity and nonlinearity in the IR. After careful consideration, we have chosen $\beta_x^* = 60$ mm, which strikes a balance between beam-beam effects and nonlinear dynamics. This value is a revision from earlier designs that used 90 mm and 40 mm, reflecting our improved understanding of the trade-offs involved.

Table 1: Comparison of IR key design parameters among the new-generation e+/e- Colliders

| Parameters | $2\theta_c$ (mrad) | $L^*$ (m) | $\beta_y^*$ (mm) | $\beta_x^*$ (mm) | Chrom. $W_y = L^*/\beta_y^*$ |
|---|---|---|---|---|---|
| SuperKEKB (LER/HER) | 83 | 0.835/1.41 | 0.27/0.30 | 32/25 | 3462/2352 |
| SuperB (LER/HER) | 66 | 0.6 | 0.21/0.25 | 32/26 | 2857/2400 |
| CEPC(Z) | 33 | 1.9 | 1 | 300 | 1900 |
| FCC-ee(Z) | 30 | 2.2 | 0.7 | 110 | 3143 |
| BINP-SCTF | 60 | 0.9 | 1 | 100 | 900 |
| STCF | 60 | 0.9 | 0.8 | 60 | 1125 |

In summary, the selected parameters represent a carefully balanced set that optimizes the performance of the IR while respecting the physical and engineering constraints. The following sections will demonstrate how these choices are embodied in the optics design and how they contribute to the overall performance.

## 3. Modular lattice design and nonlinear performance analysis of the STCF IR
### 3.1 Evolution of IR lattice design in circular e+/e- colliders

The core goal of IR optics design is to achieve a low beta function at the IP for high luminosity while ensuring a sufficiently large dynamic aperture. For new-generation circular e+/e- colliders, beam lifetime is substantially shortened—with Touschek lifetime dominating in the medium-low energy range, and Beamstrahlung lifetime in the high-energy range [25]—making the achievement of a sufficiently large momentum aperture an essential optimization target.

For first-generation single-ring circular e+/e- colliders such as SPEAR, DORIS, PETRA, BEPC, and PEP, $\beta_y^*$ was typically on the order of 10 cm or larger, achieved using FF quadrupole doublets in their IR systems. Given the relatively small natural chromaticity of these colliders, global chromaticity compensation was universally adopted—relying on sextupoles placed in the ring arcs to correct chromaticity.

As $\beta_y^*$ was pushed to smaller values, the induced natural chromaticity (and even high-order chromaticity) increased significantly, gradually becoming the primary factor limiting the collider ring's momentum acceptance. To address this, second-generation B-factory colliders (e.g., KEKB, PEP-II)—which adopted a double-ring, small-angle crossing scheme and achieved $\beta_y^*$ of 1–2 cm—introduced local chromaticity correction. This strategy originated from the final focus system of the Stanford Linear Collider (SLC) [26], which features a specially designed "small arc" with dispersion in the IR to place a sextupole pair separated by a –I transformation, enabling chromatic correction while canceling all geometric aberrations. However, design differences emerged between the two B-factories: PEP-II implemented separate horizontal and vertical local chromaticity correction sections (CCX, CCY) within the IR [27]; whereas KEKB only installed vertical local chromaticity correction (CCY) in the IR, with the remaining horizontal chromaticity corrected by the sextupole pair in the arc cell closest to the IR [28].

In new-generation crab-waist colliders, the extremely low $\beta_y^*$ (millimeter scale or even sub-millimeter) demands more powerful local chromaticity correction sections due to the induced larger chromaticity. Moreover, specialized IR optics must be designed to accommodate crab sextupoles, which require specific phase advances from the IP and tailored strengths. Notably, the phase advance between the crab sextupoles and the IP, as well as the crab sextupole strength, are derived for on-momentum particles using a linear transfer map (see Ref. [4] or the Appendix). For off-momentum particles, the transfer map between the crab sextupole and IP varies with momentum deviation $\delta$ (=$\Delta p/p_0$), altering the Twiss functions $\beta/\alpha$ at the IP and CS, and the phase advances between CS and IP. This introduces undesirable nonlinear terms that ultimately limit the off-momentum dynamic aperture (DA) and momentum aperture (MA). To mitigate this, crab sextupoles are typically placed in nearly achromatic regions—formed by separate vertical and horizontal local chromaticity correction sections. This explains why the IR optics of SuperB [7], CEPC [5], FCC-ee (designed by P. Raimondi [29]) and BINP-SCTF [30] all locate crab sextupoles at the outer ends of CCY and CCX in the IR.

A. Bogomyagkov previously proposed placing the crab sextupole closer to the IP for the FCC-ee IR [12]; this approach was later applied to the BINP-SCTF IR [31], yielding better on-momentum DA but significantly reducing 6D DA when crab sextupoles were activated. Consequently, BINP-SCTF ultimately also adopted the layout with crab sextupoles outside the separated chromatic correction sections [29]. Notably, the FCC-ee IR designed by K. Oide [32] incorporates only CCY within the IR and integrates the crab sextupole into the second one of the CCY sextupole pair—where dispersion is suppressed. This yields a more compact IR but requires stronger CCY sextupoles and leaves horizontal chromaticity between the crab sextupole pair uncorrected, potentially impairing off-momentum dynamics performance. Due to spatial and geometric constraints, SuperKEKB's IR optics adopted Oide's scheme, utilizing the second one in the CCY sextupole pair as the crab sextupole while counteracting nonlinearities from the first CCY sextupole—requiring the second sextupole to be stronger than the first. However, residual dispersion at the crab sextupole location may still limit the off-momentum DA [15].

**3.2 Modular optics design and key features of the STCF IR**

Inspired by the IR optics layouts of new-generation e+/e- colliders, the STCF IR lattice adopts a modular functional design. The design strategy aims to minimize nonlinear and off-momentum dynamic effects during the linear lattice design by incorporating phase-advance-based nonlinear cancellation for the main chromatic sextupole pairs, reducing the adverse impact of sextupoles (like additional sextupoles for high-order chromaticity correction and crab sextupoles), and minimizing the dispersion invariant $H_x$ along the IR to improve local momentum aperture (LMA) [33].

Figure 1 presents the STCF IR layout for both rings and the modular configuration of a half-IR. The half-IR comprises: the final focus telescope section (FFT), CCY, CCX, crab sextupole section (CS), and several matching sections (MCY, YMX, XMC, and MS).

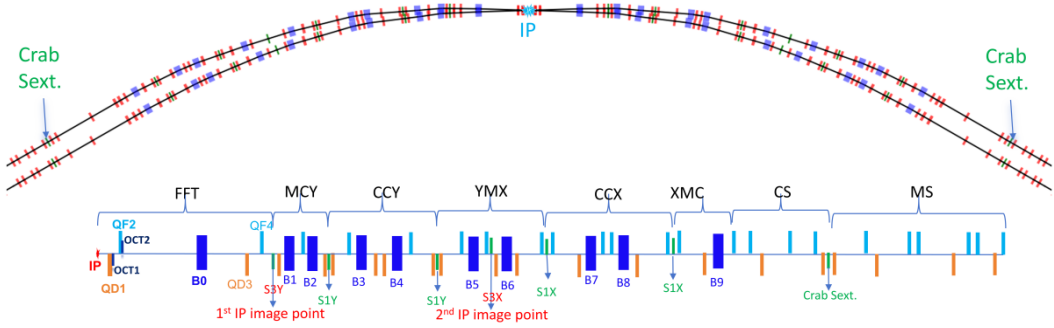

Figure 1: STCF IR layout.

Key optical features of each module are summarized below:

- FFT section: This module compresses the $\beta$-function at the IP using two quadrupole doublets—one superconducting and one normal-conducting. The superconducting doublet achieves an ultra-low $\beta^*$ at the IP, while the normal-conducting doublet forms the first IP mirror point with $\alpha_x=\alpha_y=0$ and phase advances to IP $\mu_x=\mu_y=\pi$ [34]. This configuration allows IP optics tuning by adjusting the mirror-point $\beta$-functions without modifying internal components. A weak bending magnet (B0) is inserted between the two doublets to generate local dispersion at the first IP image point (facilitating the sextupole installation for high-order chromaticity correction) and enhance dispersion at the CCY sextupoles to improve chromatic correction efficiency. Particularly, B0 is the weakest magnet in the IR and the entire collider ring with bending angle of 1° and is positioned 8.5 m from the IP—this deliberate placement reduces synchrotron radiation background in the detector, addressing one MDI challenge.

- FFT-to-CCY matching section (MCY): This section employs a FODO-like cell (starting with a defocusing quadrupole) to ensure the required Twiss parameters ($\beta_y \gg \beta_x$ and $\alpha_y = 0$) at the vertical chromaticity correction sextupole (S1Y). A key design here is that the vertical phase advance from S1Y to the first FF quadrupole (QD1) is approximately $\pi$ [35] and can be fine-tuned for second-order chromaticity correction. Bending magnets B1 and B2 are inserted in the MCY to enhance dispersion at S1Y, reducing the required strength of S1Y and thus mitigating associated nonlinearities, and help control $H_x$.

- CCY section: This module implements an exact -I transformation between the S1Y sextupole pair to cancel sextupole-induced nonlinearities. It is noted that the general conditions for canceling geometric aberrations in sextupole pairs are detailed in Ref. [36], while the -I map is a special case.

The CCY section achieves this -I map via two mirror-symmetric FODO-like cells, each designed with π/2 phase advances (for both *x* and *y* planes) and $\alpha_x=\alpha_y=0$ at the section midpoint. The drift lengths within this section are carefully tuned to achieve the exact -I condition. Dipoles B3 and B4 are installed to create symmetric dispersion distribution across the CCY and further refine $H_x$ control.

- CCY-to-CCX matching section (YMX): This section transitions Twiss parameters from $\beta_y \gg \beta_x$ at S1Y to $\beta_x \gg \beta_y$ at the first horizontal chromaticity correction sextupole (S1X). The horizontal phase advance from S1X to the second FF quadrupole (QF2) is set near $3\pi$ to support second-order chromatic correction. Besides, a second IP mirror point is formed with $\alpha_x = 0$. Dipoles B5 and B6 are added to enhance dispersion at S1X (lowering sextupole strength demands) and maintain $H_x$ control.

- CCX section: Similar to the CCY section, the CCX implements a -I transformation between the S1X sextupole pair, with dipoles B7 and B8 providing symmetric dispersion while preserving $H_x$ control.

- Dispersion Suppression section (XMC): Dispersion is suppressed at the crab sextupole location using dipole B9.

- CS section: This section utilizes six quadrupoles to meet the phase advance constraints from the crab sextupole to the IP as $\mu_x = 6\pi, \mu_y = 5.5\pi$ while keeping $\alpha_x = 0$, $\alpha_y = 0$ at the crab sextupole. Another optics design feature is minimizing CS strength by prioritizing $\beta_y \gg \beta_x$, and requiring $\beta_x$ as small as possible but larger than 1—this configuration mitigates undesired crab sextupole nonlinear effects, with details in Appendix. In the present design, $\beta_y$ = 900 m and $\beta_x$ = 2.8 m are adopted, which gives the integral gradient field of crab sextupole about 3.39 m$^{-2}$.

- Matching Section (MS): This section deploys several quadrupoles to match the $\beta/\alpha$ functions to the long straight section while adjusting phase advances between the IR and the adjacent arc.

The linear optics functions of the STCF IR are shown in Fig. 2. All dipoles are standardized to 1m length to reduce manufacturing cost, and their strengths are optimized to limit $H_x$ (except B0, fixed at 1°). The locations of dipoles ensure a 1.5–2 m separation between the two rings. The total bending angle over the IR is 60°, driven by the collider ring's geometric layout. To achieve the 60 mrad IP crossing angle, an asymmetric 30 mrad split is adopted: the outer-ring half-IR bends 30 mrad more, while the inner-ring half-IR bends 30 mrad less. Though this creates asymmetric dispersion across the IR, the design maintains a global $H_x \leq 0.02$ m—a critical optimization for enhancing LMA.

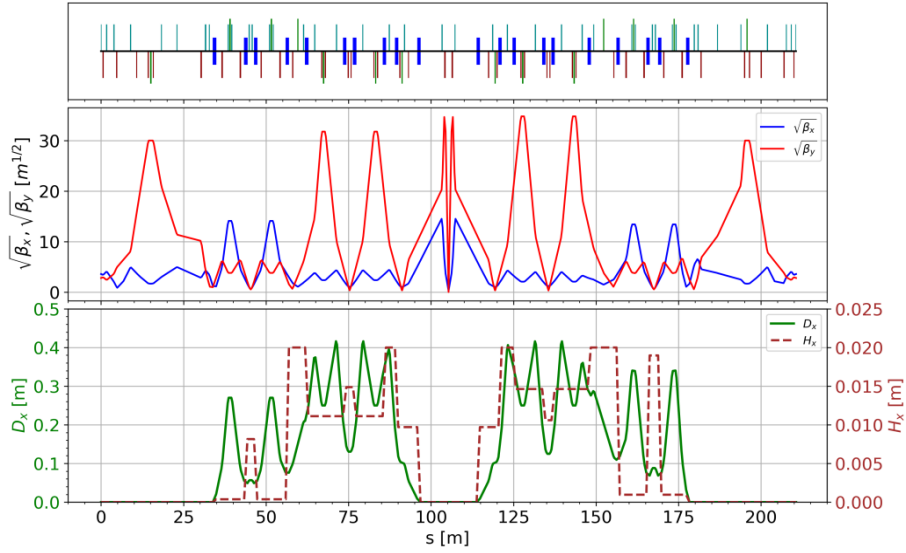

Figure 2: Linear optics functions for the STCF IR.

### 3.3 Local chromaticity correction scheme

To reduce beam loss and improve injection efficiency, the STCF aims for a beam lifetime (predominantly governed by the Touschek lifetime) exceeding 200 s. To achieve the target Touschek lifetime for STCF, the momentum acceptance should be expanded to approximately ±1.5%. Besides, to mitigate the adverse impact of crab sextupoles on the momentum aperture, the IR optics should exhibit minimal chromatic dependence at the crab sextupole locations. These requirements necessitate a systematic high-order local chromaticity correction strategy for the IR optics, extending to the third order [37].

The correction methodology is implemented as follows:

- First-order vertical and horizontal chromaticity is corrected using the main sextupole pairs S1Y and S1X, respectively, which are placed in an exact -I transformation configuration to cancel geometric aberrations while compensating for linear chromaticity.
- Second-order chromaticity is suppressed by carefully optimizing the phase advances from the S1Y and S1X sextupoles to the FF quadrupoles QD1 and QF2, respectively, as described in section 3.2. This is achieved through fine-tuning of quadrupole strengths in the MCY and YMX matching sections without introducing additional nonlinear elements. Proper phase advance selection between the FF quadrupoles and the sextupole pairs also helps minimize the leakage of the leakage of the chromatic beta function amplitude induced by FF quadrupoles into the ring arcs [38].
- Third-order chromaticity is compensated using additional sextupoles (S3Y and S3X), commonly referred to as Brinkmann sextupoles [39], positioned at the first and second IP image points where the beta functions reach local minima. This optics configuration enhances their effectiveness in correcting higher-order chromatic effects while limiting nonlinear side effects.

This systematic approach to local chromaticity correction is well-aligned with strategies adopted in other new-generation e+/e- colliders [5,7,9,15,29,30]. After implementing third-order correction in the STCF IR, the off-momentum optical behavior—described by the Montague $W$ functions [40]—is well controlled at both the IP and the crab sextupole locations, as shown in Fig. 3. Notably, the phase advances from S1Y and S1X to the IP are also optimized to minimize the $W$ functions at the crab sextupole locations. A sharp decline in $W_y$ at S1Y and $W_x$ at S1X in Fig. 3 indicates strong decoupling between horizontal and vertical planes, underscoring the orthogonality of the corrected optics.

Nevertheless, a residual second-order dispersion $D'_x$ outside the IR may still influence off-momentum dynamics. In addition, the residual second-order vertical chromaticity remains non-negligible, which is displayed in Fig.6 (right), ultimately limiting the achievable momentum acceptance. These observations highlight areas for further refinement in future optimizations.

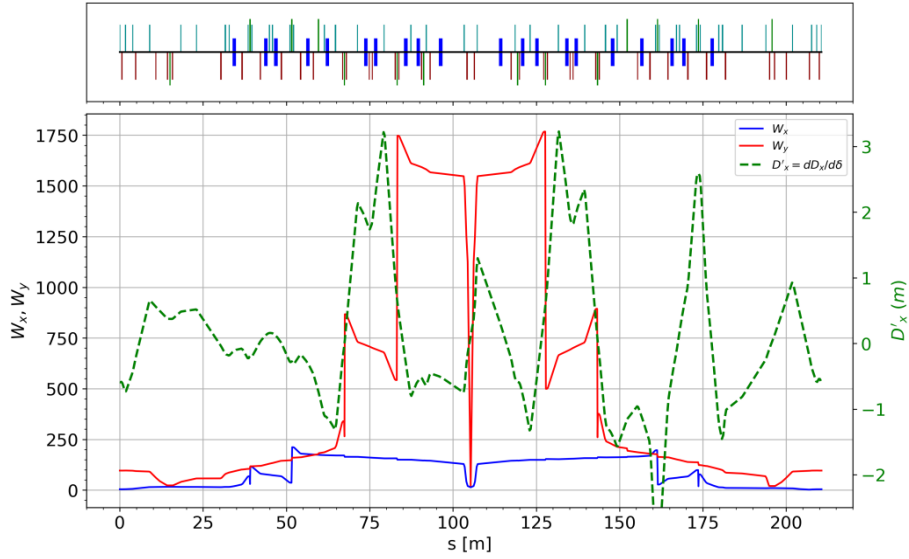

Figure 3: Montague functions and second-order dispersion in the STCF IR with third-order chromaticity correction.

### 3.4 Nonlinear dynamics analysis via resonance driving terms

To assess the effectiveness of the local chromaticity correction in the STCF IR, we computed the distribution of resonance driving terms (RDTs) [41] along the IR using E. Forest's FPP/PTC code, as shown in Fig. 4.

The results indicate that first-order geometric terms are fully canceled outside the crab sextupole pairs, while transverse coupling terms such as $h_{10110}, h_{10200}$, and $h_{10020}$ persist at the IP. This is consistent with the intentional crab-waist effect generated by the $xy^2$ Hamiltonian component of the crab sextupole [42]. Thanks to the exact -I transformation applied between the CCY and CCX sextupole pairs, amplitude-dependent tune shifts (ADTs) and second-order geometric terms are also largely canceled outside these sections.

It is worth noting that a sextupole pair separated by a -I transformation can, in principle, compensate for geometric aberrations—both second-order and higher-order—introduced during chromaticity correction [34]. However, this result is strictly valid only under the thin-lens approximation and for on-momentum particles. When accounting for the finite length of the sextupoles or off-momentum beam conditions, the ideal -I transformation is disrupted, leading to residual nonlinearities. For instance, certain RDTs such as $h_{00220}$ and $h_{00310}$ are not completely canceled outside the CCY sextupole pair due to finite-length effects, though the magnitude of these residuals remains small. This suggests that, in the case of STCF, the impact of finite sextupole length is not severe.

Nonetheless, in the off-momentum case, chromatic-geometric terms and chromatic ADTs remain only partially compensated due to the breakdown of the -I condition in both CCY and CCX sections. This results in relatively large residual driving terms at the IR exit, potentially limiting the off-momentum dynamic aperture. This motivates us to perform the global nonlinear optimization for the collider ring with crab sextupoles activated when the IR is integrated into the ring, see Section 4.1. Besides, this also highlights an important area for future local optimization of chromatic RDTs within the IR.

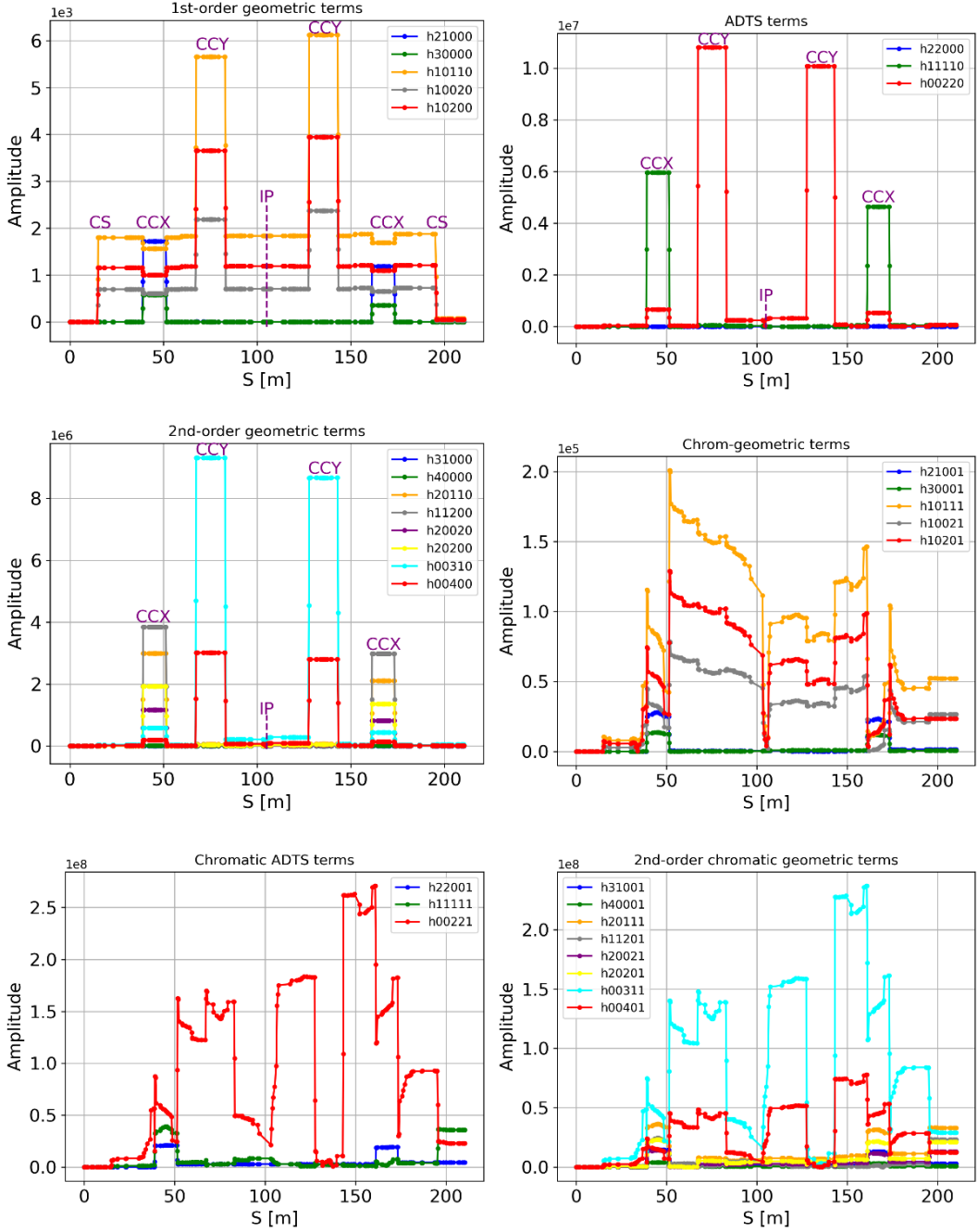

Figure 4: Distribution of resonance driving terms (RDTs) across the STCF IR.

## 4. Performance evaluation of the STCF IR integrated into the collider ring
### 4.1 Global dynamics performance of the collider ring with the integrated IR

To assess the IR performance under realistic operating conditions, the designed IR optics was integrated into the full STCF collider ring lattice. Comprehensive details on the collider ring—including the ring layout, arc optics, global optimization procedure, and dynamics performance—are provided in Ref. [20]; here, we summarize only the aspects directly relevant to the IR performance analysis.

The collider ring features a quasi-two-fold symmetric structure comprising one IR, four 60° arcs, two 30° arcs, and multiple straight sections for injection, extraction, damping wigglers, RF cavities, and collimation. The arcs employ standard 90° FODO cells—replacing the earlier HMBA design [19]—to achieve a target emittance of ~5 nm·rad while improving the momentum compaction factor, minimizing the H-invariant, and enabling more flexible sextupole configuration. Chromaticity outside the IR is corrected using an interleaved -I transformation sextupole scheme, which reduces individual magnet strength while canceling first-order geometric aberrations, thereby ensuring the IR's designed performance is not masked by arc-induced nonlinearities. A total of 48 sextupole families (8 in the IR, 40 in the arcs) were tuned using the PAMKIT code [43] with multi-objective algorithms for global nonlinear optimization. During optimization, crab sextupoles were always enabled and IR chromatic sextupole strengths were permitted slight variations from their nominal design values.

Figure 5 shows the variation of the Montague function and second-order dispersion along the ring after optimization. The Montague function remains well-controlled within the IR, while second-order dispersion exhibits relatively large perturbations globally primarily due to the leakage of residual second-order dispersion in the IR into arcs. Figure 6 presents the tune shift with momentum for the full ring (left) and the variation of phase advance for the IR only (right), demonstrating a momentum acceptance of approximately ±1.5%—nearing the target required for the STCF's Touschek lifetime. Besides, the close resemblance of full-ring tune shift to the IR-only phase shift indicates that the momentum acceptance is primarily limited by residual vertical second-order and horizontal third-order chromaticity originating in the IR.

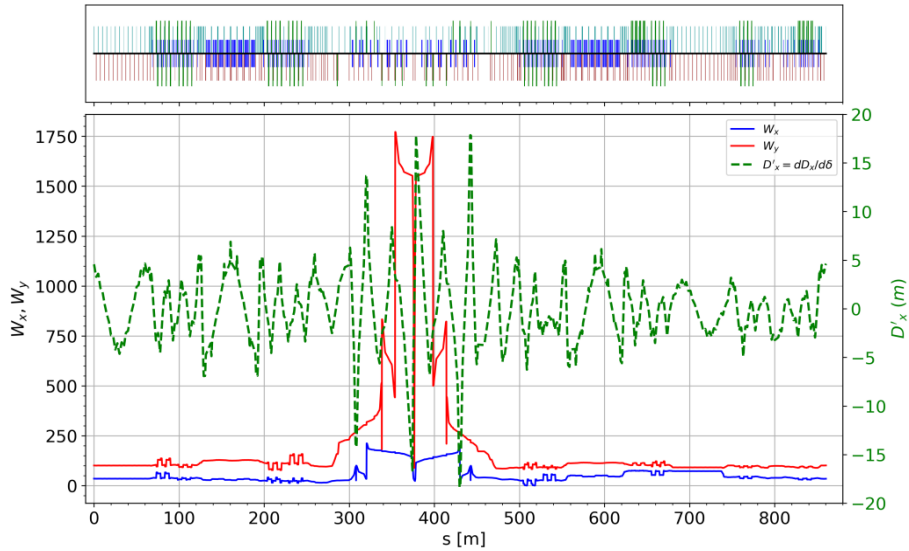

Figure 5: Montague function and the second-order dispersion along the ring.

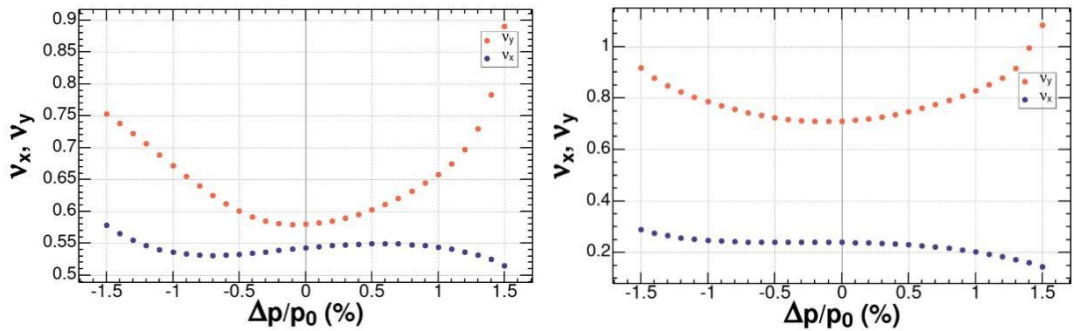

Fig. 6: The fractional tune variation with the momentum deviation for the ring (left) and the fractional phase advance variation for IR only (right).

The 6D DA was simulated using the SAD code [44] for both CS-on and CS-off scenarios, including synchrotron motion, synchrotron radiation, quantum fluctuations, tapering, and high-order kinematic terms (depending on high powers of transverse momentum [12]). For this baseline assessment, only nonlinearities from sextupoles (chromatic and crab sextupoles) and high-order kinematic terms are involved. As shown in Fig. 7, activating the crab sextupoles causes only a slight reduction in both on-momentum and off-momentum DA, indicating that the current IR is well optimized. This is attributed to specific IR optimizations: 1) Linear optics design at the CS minimizes undesired Hamiltonian terms for on-momentum particles; 2) Local chromaticity correction up to the third-order in the IR reduces the variation of off-momentum Twiss parameters at the CS; 3) Nonlinearities from chromatic sextupole pairs in the IR is well counteracted by achieving -I transformation. Touschek lifetime with CS on can reach 350 s at $0.94 \times 10^{35}$ cm$^{-2}$s$^{-1}$ and 2 GeV, satisfying the STCF design goal. The dynamics performance of the collider ring integrating more realistic IR that includes nonlinearities from Maxwellian fringe field and detector solenoid will be addressed in Section 4.3 and 4.4, respectively.

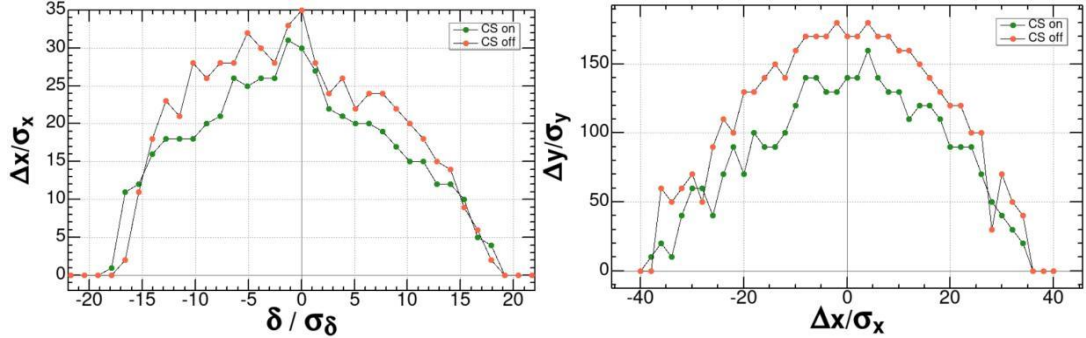

Figure 7: Off-momentum and on-momentum dynamic apertures for the STCF ring.

RDTs were also computed to assess the nonlinear and chromatic behaviour across the full ring, see Fig. 8. The results demonstrate that the nonlinear contribution from the arcs to the RDTs remain quite small, while the IR is the dominant source of nonlinearities, underscoring the importance of careful local chromaticity correction and nonlinear suppression in the IR. Besides, crab sextupoles can induce first-order and second-order chrom-geometric terms and chromatic ADTs for off-momentum particles, see Fig. 4 and Fig. 8. This also explains why we perform the global nonlinear optimization for the collider ring with crab sextupoles activated.

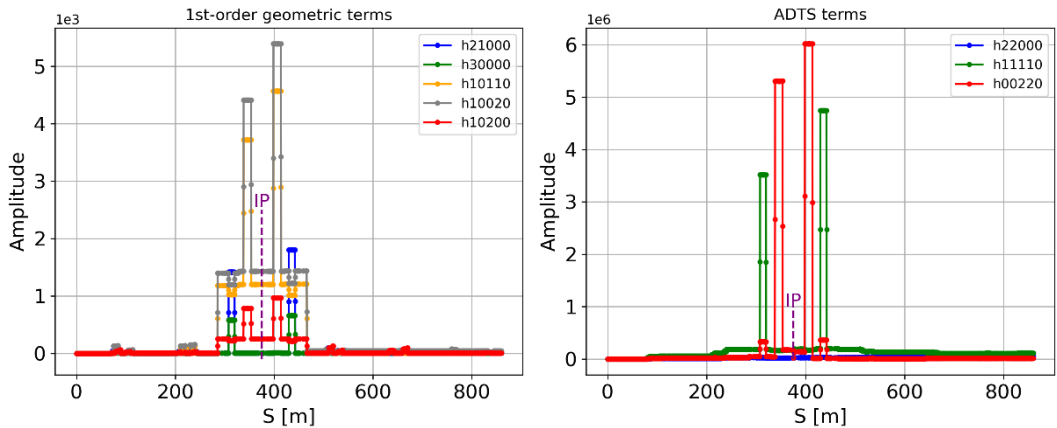

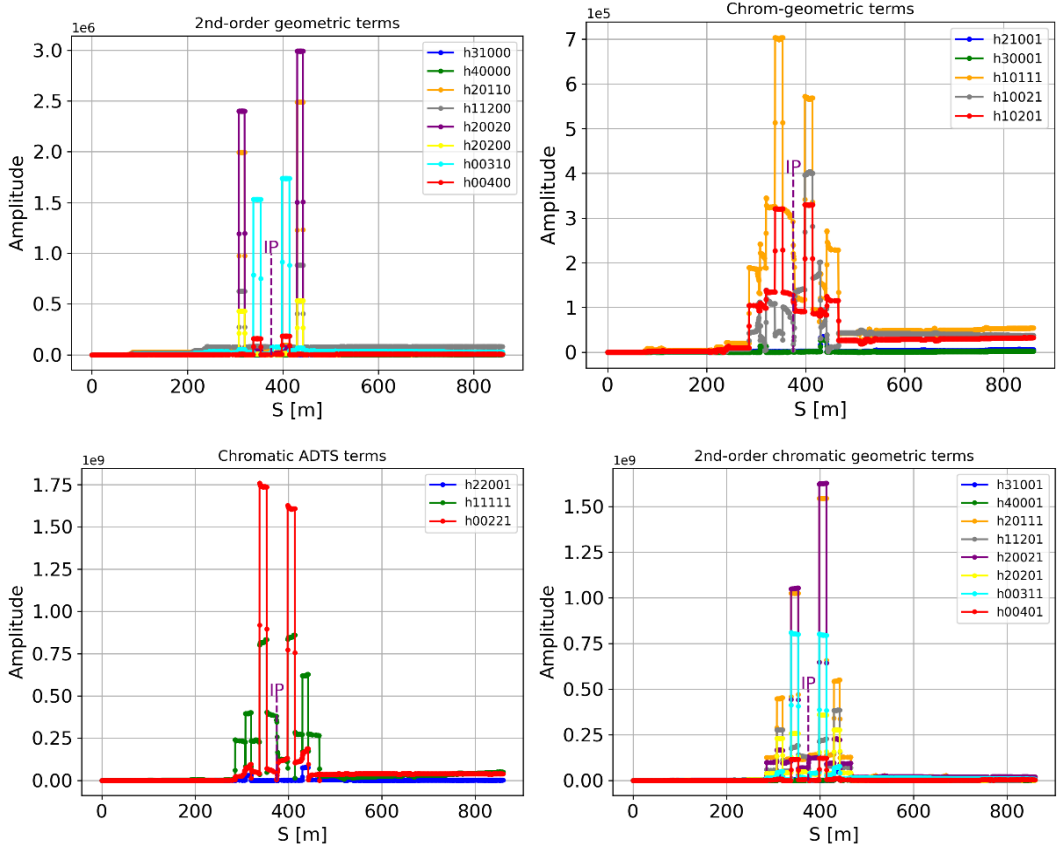

Figure 8: RDTs distribution across the STCF ring.

**4.2 Validation of the crab-waist effect for the integrated IR**

Given that the crab sextupoles do not significantly degrade the dynamic aperture, it is essential to verify whether they function as intended. The effectiveness of the crab-waist scheme is confirmed through the following analyses:

(1) Hamiltonian and RDT Analysis

The crab sextupole introduces an $xy^2$ term in the Hamiltonian, which shifts the vertical beta waist at the IP along the central axis of the opposing beam as a function of horizontal amplitude. This $xy^2$ component can generates specific transverse RDTs $h_{10110}$, $h_{10200}$, and $h_{10020}$[]. As shown in Fig. 8 (variation of first-order geometric terms along the ring), these terms propagate to the IP and remain non-zero there. This observation aligns with the RDT analysis in Section 3.4 and confirms that the crab sextupoles can effectively introduce the intended crab-waist effects at the IP.

(2) Beam-Beam Simulation with Lattice

Beam-beam simulations (incorporating the full collider ring lattice and realistic beam parameters, e.g., bunch charge, emittance) represent the most direct and rigorous approach to validating the crab-waist effect, as luminosity is the ultimate performance metric the effect aims to enhance. The simulation results [45] demonstrate the luminosity with CS on is significantly higher than with CS off. This gain arises because the CS suppress the detrimental betatron beam-beam resonances that would otherwise broaden the beam tune spread and increase the beam envelope.

(3) Single-Pass Particle Tracking

Multi-particle tracking offers direct visualization of the crab-waist effect via the beam density distribution at the IP. Figure 9 displays the beam density distribution at the IP after single-pass tracking, with two distinct scenarios: 1) With CS off: The z-x plane density distribution is normal, indicating no vertical waist shift; 2) With CS on: A clear density modulation emerges in the z-x plane, which directly reflects the vertical $\beta_y$ waist shift induced by the crab sextupole.

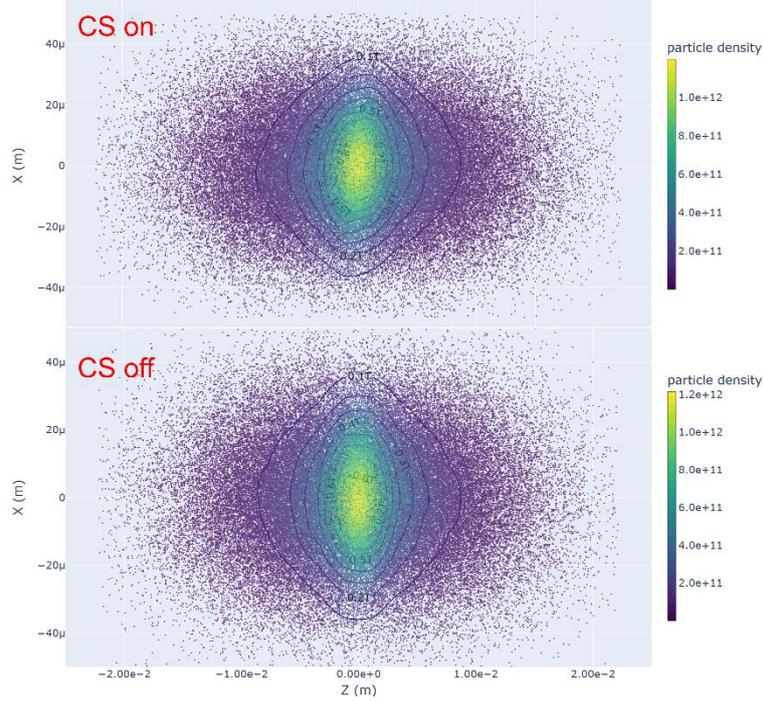

Figure 9: Particle density distribution at the IP after tracking with crab sextupole on and off.

**4.3 IR fringe field compensation**

Fringe fields exist in all magnets, introducing additional nonlinearities that degrade the dynamic aperture. Unlike high-energy colliders like FCC-ee, where fringe field effects are the relatively minor [32], the fringe field effects in the STCF are far more critical, primarily due to its lower energy regime. To quantify these impacts, we performed simulations using the SAD code, incorporating Maxwellian fringe field effects from dipoles, quadrupoles, and sextupoles in the STCF collider ring. The results indicate that fringe fields from the ultra-strong FF superconducting quadrupoles are the dominant source of DA degradation, as illustrated in Fig. 10.

To mitigate this, octupole coils have been installed at the IP-opposite ends of the FF quadrupoles (QD1 and QF2, see Fig. 1). This compensation strategy is motivated by the similarity between the Hamiltonian of a quadrupole fringe field and that of a thin octupole magnet [46], allowing partial compensation of the nonlinear fringe components. As shown in Fig. 10, the introduction of octupole compensation leads to a clear recovery in DA performance, directly validating the effectiveness of this correction strategy.

Particularly, the DA drops sharply with CS on compared to CS off when fringe fields from the FF quadrupoles are included, see Fig. 10. This observation reveals that the interference between CS and other IR nonlinearities can limit the DA. To address this, the strengths of the compensation octupoles were optimized as part of the global nonlinear optimization process with CS activated, ensuring the compensation scheme aligns with realistic operating conditions and maximizes DA.

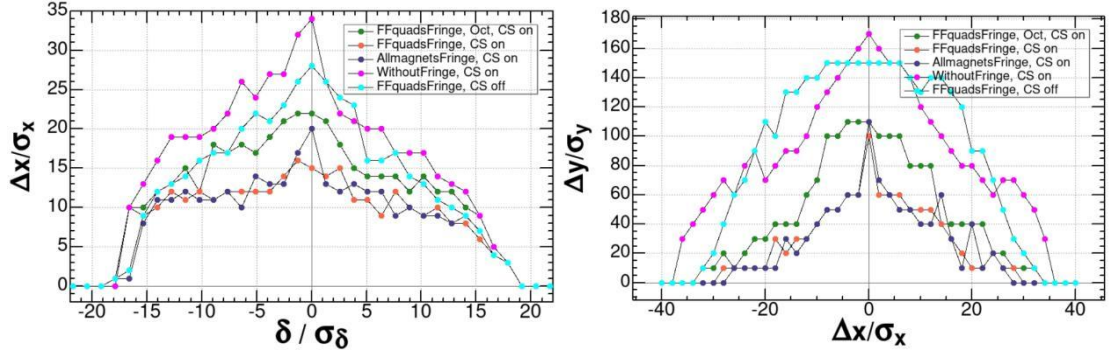

Figure 10: Off-momentum and on-momentum dynamic apertures considering the fringe field effects from different magnets and octupole compensation with CS on for the STCF ring. In comparison, the DA is also shown considering FF quadrupoles fringe fields with CS off.

**4.4 Detector solenoid field compensation**

The detector solenoid field, if not properly compensated, can significantly increase vertical emittance and lead to substantial luminosity loss. This occurs primarily through two mechanisms: (1) the longitudinal $B_z$ component introduces horizontal–vertical coupling, and (2) the radial fringe field $B_r$ generates vertical dispersion due to the crossing angle. To suppress these effects, a local perfect compensation scheme using anti-solenoids is implemented for STCF. This scheme requires the integrated $B_z$ field to cancel completely between the IP and the pole face of the final quadrupole QD1, with the $B_z$ field vanishing inside the superconducting quadrupoles QD1 and QF2.

In this study, a step-function solenoid model based on a configuration similar to FCC-ee [31] is adopted for evaluation: a detector solenoid field of -1 T extends from -0.5 m to +0.5 m around the IP, while an anti-solenoid field of -1.25 T covers ±0.5 m to ±0.9 m. Detailed engineering integration of the solenoid compensation system will be addressed in Section 5.2.

As shown in Fig.11, with perfect local compensation, both the vertical dispersion and vertical closed orbit are effectively confined near the IP, with no significant leakage into outside regions. The apparent sharp change in the horizontal closed orbit observed in Fig. 11 is not physical, but arises from the coordinate system transition from the beam axis to the solenoid central axis in the SAD simulation when the beam enters the solenoid field. Horizontal–vertical coupling is also eliminated due to the zero net integrated $B_z$ field. The resulting vertical emittance growth from the combined effect of the compensation solenoid fringe field and the crossing angle is less than 1 pm. This is negligible compared to the STCF ring's design vertical emittance of about 50 pm, ensuring almost no luminosity loss. This margin is considerably more relaxed than in FCC-ee or CEPC, where the 1-2 pm design vertical emittance imposes far stricter constraints on solenoid-induced emittance growth [47]. The STCF's larger design emittance provides inherent tolerance, making the local perfect compensation scheme even more effective.

The weak optics focusing effect introduced by the solenoids is corrected by fine-tuning the strengths of FF quadrupoles to restore the ring tunes. This adjustment has negligible impact on the beta functions and introduces only minor, acceptable perturbations in the horizontal dispersion. Furthermore, as illustrated in Fig. 12, the dynamic aperture is fully recovered under the perfect solenoid compensation scheme. These results confirm that the fully local anti-solenoid compensation approach effectively eliminates the solenoid's adverse effects without introducing new nonlinearities and thus is well-suited to the STCF collider.

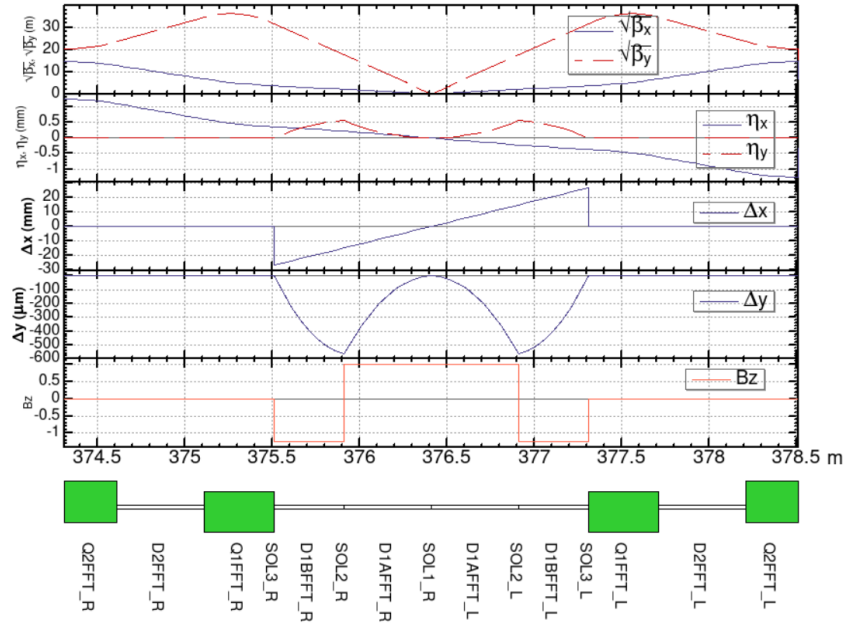

Figure 11: The influence of the local perfect anti-solenoid compensation scheme on the beta functions, dispersions, horizontal closed orbit and vertical closed orbit. The solenoid field is displayed at the bottom.

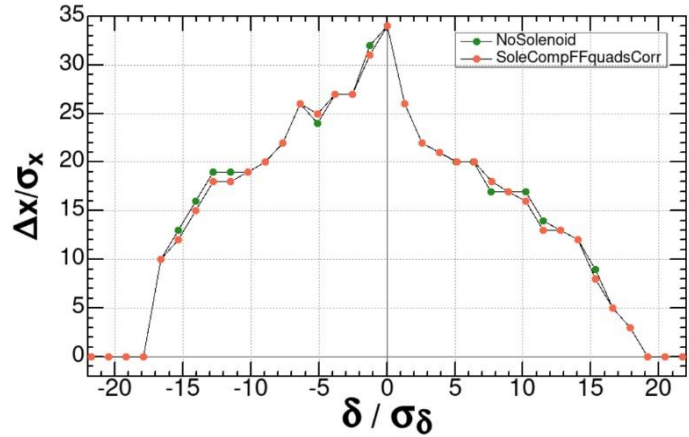

Figure 12: The influence of the local perfect anti-solenoid compensation scheme on dynamic aperture.

## 5. Physical design of the machine-detector interface for the STCF
### 5.1 MDI design requirements and constraints

The MDI serves as the critical junction between the collider rings and the detector spectrometer, essential to both accelerator operation and physics data acquisition. In the STCF, the electron and positron beams are tightly focused by FF superconducting quadrupoles near the IP to achieve high luminosity. This strong focusing requires the detector's inner tracking system to be positioned extremely close to the IP, resulting in highly limited space for accelerator elements. The MDI design is further challenged by the collider's ultra-high luminosity, large crossing angle, strong focusing requirements, and short beam lifetime.

The MDI region must integrate a wide range of accelerator systems—including lattice design,

beam dynamics, superconducting magnets, diagnostics, vacuum, collimation, and alignment—as well as detector subsystems such as background simulation, inner detector layout, and luminosity monitoring. Achieving spatial and functional compatibility between the accelerator and detector requires close interdisciplinary collaboration, with the overarching goal of maximizing luminosity and operational stability while minimizing experimental background.

This section presents the physics-related design aspects of the STCF MDI, covering the mechanical layout, synchrotron radiation management, beam-stay-clear specifications, and preliminary background estimates.

**5.2 MDI layout and integration**

The layout and longitudinal positions of accelerator components in the MDI are fundamentally determined by the IR lattice design (see Section 3.2). A key constraint driving the MDI layout is the detector's requirement to maximize the solid angle for physics experiments—this limits accelerator elements in the MDI to a 15° conical region with the IP as the apex (a critical spatial boundary to avoid blocking the detector's sensitive acceptance).

Figure 13 presents a two-dimensional layout of the STCF half-MDI, showing the positions and dimensions of components within ±3.5 m of the IP. These include detector boundaries, central beam pipe, vacuum pipe structures, superconducting magnets and cryostat, beam instruments, and mechanical supports.

The cryostat integrates two superconducting quadrupoles (QD1/QF2), two compensating solenoids, a screening solenoid, corrector magnets, two beam position monitors (BPMs), and tungsten shielding. The compensating solenoid located IP-upstream of QD1 cancels the integrated longitudinal magnetic field of the detector solenoid between the IP and the front-end face of QD1. The screening solenoid, wound around the superconducting quadrupoles, ensures a zero $B_z$ field within the quadrupole regions. A second compensating solenoid located QF2-upstream counteracts the long-range fringe field of the detector solenoid.

The central beam pipe comprises an IP section and a transition section. The IP beam pipe is constructed of beryllium, with an inner gold coating to mitigate unpredictable synchrotron radiation background into detector (see Section 5.3) and reduce beam impedance. It has a length of 215 mm (half-length 107.5 mm) and an inner diameter of 30 mm, matching the inner diameter of the beam pipe at QD1 to avoid forming a cavity structure at the IP pipe (which would otherwise cause high-order mode (HOM) power deposition). The transition pipe, currently specified as tantalum with a copper-coated interior, expands from 110 mm to 500 mm in length, transitioning from a 30 mm round profile to a racetrack-shaped cross-section measuring 60 mm wide and 30 mm high, which accommodates the transition to dual-aperture beam pipes.

The Y chamber (or remote vacuum connector, RVC) is a key junction where the single-aperture central beam pipe splits into two 30 mm inner-diameter round channels (for e+ and e- beams), extending to the end of QD1. Beyond QD1, the pipes transition over 0.5 m to 50 mm inner-diameter at the entrance to QF2, and further expands to 67 mm within another 0.5 m section. Critical design considerations here include:

- Smooth transitions: The pipe transitions (single-to-dual aperture and diameter changes) are designed to be as smooth as possible—this minimizes beam impedance;
- Diagnostic integration: An eight-button BPM is installed before the bellows at the RVC, offering potential for real-time beam position monitoring near the IP, a capability not yet realized in existing colliders.

CST simulations of electromagnetic modes indicate a total beam power loss of approximately 587 W for both beams traversing the central beam pipe and Y-chamber. About 40 W is deposited in the beryllium chamber—within its safe thermal limit—while the majority of the remaining power (dominated by capture-mode loss) occurs in the transition chamber. Ongoing optimization of the Y-chamber geometry aims to reduce capture-mode losses and further minimize impedance.

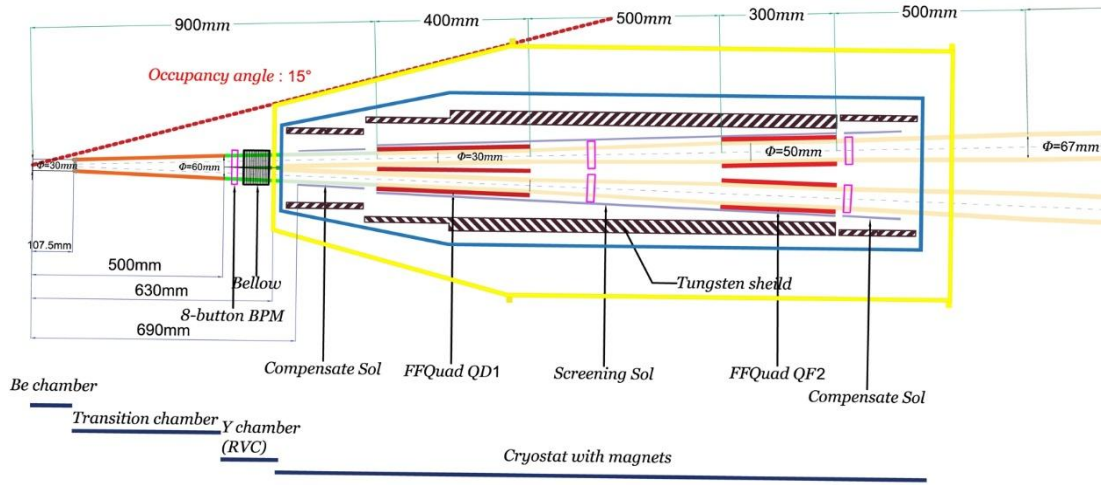

Figure 13: The 2D layout of the STCF half MDI.

**5.3 MDI synchrotron radiation**

The primary source of synchrotron radiation entering the MDI region is the upstream bending magnet B0, see Section 3.2. In the IR lattice design, B0 is intentionally positioned 8.5 m from the IP and set to a weak bending angle of only 1°—a targeted choice to minimize radiation-induced background in the MDI. The corresponding critical energy of the emitted synchrotron radiation is 0.31 keV at the beam energy of 2 GeV and 1.66 keV at 3.5 GeV. Radiation at these low critical energies is easily absorbed by the beam pipes and shielding in the MDI, and thus does not pose a major background concern.

Synchrotron radiation distributions were computed for both electron and positron beams entering the IP from the STCF's outer and inner rings. As shown in Fig. 14 (illustrated for the positron beam case, with directly irradiated areas marked in yellow), radiation from B0 does not strike the central beryllium beam pipe when the beam enters the IP from the outer ring—this protection arises from the effective shielding effect of the transition beam pipe, see Section 5.2. This outer-ring configuration is therefore preferred, and it also aligns with the STCF's beam injection layout for operational convenience [20].

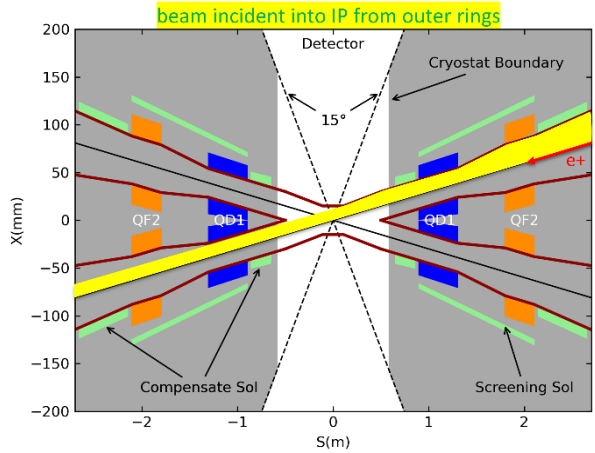

Figure 14: Synchrotron radiation fan generated by the upstream B0 within MDI for e+ beam.

Figure 15 shows the calculated line power density distribution of synchrotron radiation from B0 along the MDI beamline under beam energy of 3.5 GeV and beam current of 2 A, with power loss in different regions explicitly indicated. These results provide essential input for the thermal and mechanical design of the MDI's vacuum chambers, as well as for mask shielding (if required).

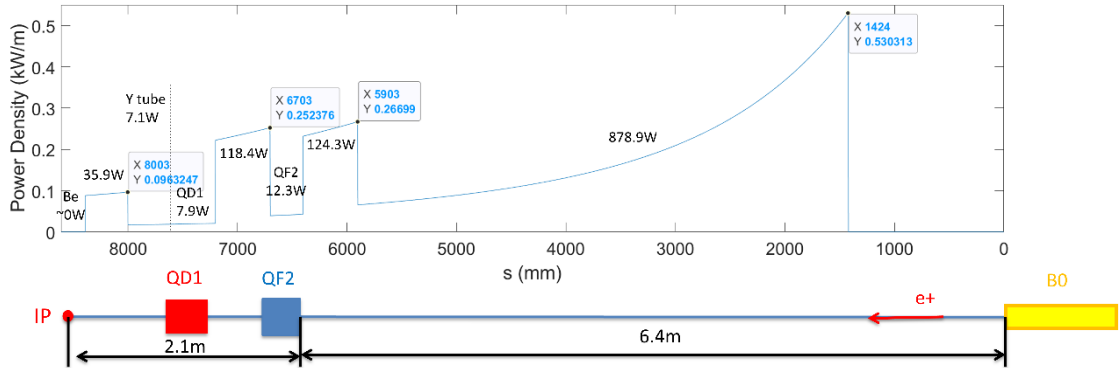

Figure 15: The line power density distribution of synchrotron radiation from B0 along the ring.

The synchrotron radiation distribution was further evaluated under extreme beam deviation scenarios: a closed orbit deviation (COD) of 1 mm and an angular deviation of 1 mrad at B0. Even under these scenarios, no synchrotron radiation hits the beryllium pipe. Additionally, radiation generated in the FF quadrupoles was analyzed and found to be entirely negligible, assuming a COD within 200 μm—this value is well above the alignment tolerance of 30 μm specified for these FF quadrupoles.

### 5.4 MDI beam stay-clear region

The definition of beam-stay-clear (BSC) regions in the MDI is essential for the mechanical and magnetic design of the FF quadrupoles. In second-generation e+/e- colliders such as PEP-II and BEPCII, the stay-clear aperture is typically defined as 15–20 times the RMS beam size plus an additional 1–2 mm to account for closed orbit deviations [48]. In contrast, new-generation colliders such as SuperKEKB, SuperB, and FCC-ee impose tighter tolerances on orbit stability (below 100 μm in the superconducting quadrupoles), allowing the BSC to be defined without explicitly including orbit errors.

For STCF, both interpretations are considered. Figure 16 illustrates the BSC boundaries at beam energies of 2 GeV and 3.5 GeV, where the dashed magenta and green lines represent the positron and electron beam envelopes, respectively. At 2 GeV, the horizontal BSC is defined as $22\sigma_x + 2$ mm (or $24\sigma_x$), and the vertical BSC as $22\sigma_y + 2$ mm (or $26\sigma_y$). At 3.5 GeV, the larger natural emittance results in a smaller BSC aperture: $9.5\sigma_x + 2$ mm (or $10.5\sigma_x$) horizontally and $9.5\sigma_y + 2$ mm (or $11.5\sigma_y$) vertically. These values assume a horizontal emittance of 5 nm at 2 GeV and 27 nm at 3.5 GeV, with the vertical emittance set to be 5% of the horizontal value—conservatively above the design coupling of 1% and 0.5% at 2 GeV and 3.5 GeV, respectively.

The relatively tight BSC at 3.5 GeV can be relaxed by increasing beta functions at the IP. For example, raising $\beta_x^*$ from 60 mm to 150 mm and $\beta_y^*$ from 0.8 mm to 1.6 mm increases the BSC to about $15\sigma_{x,y}$, with only a 24% reduction in luminosity—an acceptable trade-off for high-energy operation.

Notably, the BSC aperture at the FF superconducting quadrupoles represents the most critical region in the entire ring. This results from two primary constraints: 1) the limited physical aperture of the FF quadrupoles imposed by $L^*$ of 0.9 m and the crossing angle of 60 mrad, see Section 2; 2) the large beta functions at these positions due to extremely small $\beta_{x,y}^*$ at the IP. These conditions amplify beam sizes and enhance sensitivity to particle loss from Touschek scattering and injection processes, which in turn raises the risk of superconducting magnet quenches and increased detector background. To mitigate this risk, collimators will be placed strategically larger than 15 m from the IP (outside the MDI core) to localize particle loss away sensitive regions near the IP.

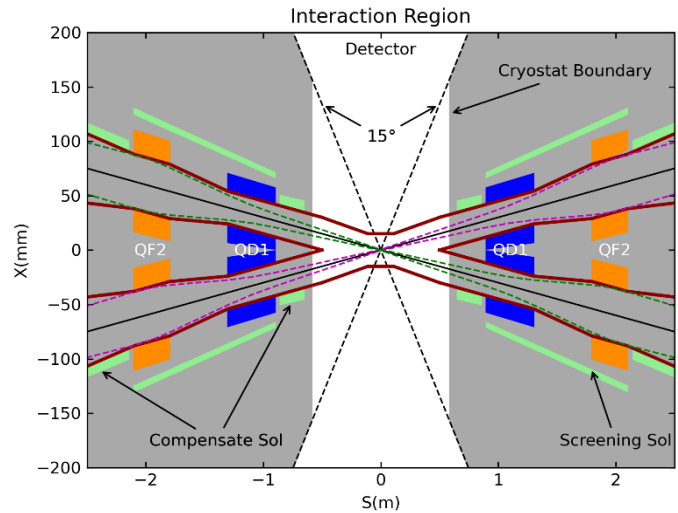

Figure 16: The beam stay-clear regions for the STCF MDI.

## 5.5 Preliminary estimation of detector background

Experimental backgrounds in the IR include beam-induced sources—such as Touschek scattering, beam-gas Coulomb scattering, beam-gas bremsstrahlung, synchrotron radiation, and injection losses—as well as luminosity-related processes like radiative Bhabha scattering and beamstrahlung [49,50].

Background evaluation for the STCF MDI involves close collaboration between accelerator and detector teams to ensure consistency between beam dynamics simulations and background assessments. The accelerator team optimizes collimator settings (located $\geq$ 15 m from the IP, outside the MDI core) to minimize beam losses from Touschek scattering and injection halo in the MDI region. The detector team subsequently quantifies backgrounds from Touschek scattering, beam-gas Coulomb scattering, and beam-gas bremsstrahlung, using the collimator configurations provided. Particle tracking is performed with the AT (Accelerator Toolbox) and SAD codes, ensuring consistency with the STCF lattice and beam parameters.

Preliminary simulations identify Touschek scattering and beam-gas Coulomb scattering as the dominant background sources in low-energy e+/e- colliders like STCF [45]. With properly placed collimators, Touschek losses in the IR are reduced to less than 1% of the total ring-wide Touschek losses—an order-of-magnitude improvement compared to the case without collimation. The same collimator settings also effectively control injection losses without compromising injection efficiency. Once Touschek scattering is mitigated, beam-gas Coulomb scattering becomes the leading background source, necessitating an ultra-high dynamic vacuum (MDI region ~$10^{-7}$ Pa, other regions $10^{-8}$ Pa) to minimize residual gas interactions.

Synchrotron radiation and beam-gas bremsstrahlung represent secondary but non-negligible backgrounds. Synchrotron radiation primarily contributes to beam pipe thermal load, though its high-energy components at 3.5 GeV may still contribute to detector background and require further assessment. Beam-gas bremsstrahlung becomes more relevant at higher energies, particularly near 3.5 GeV, further underscoring the importance of ultra-high vacuum conditions in the IR.

## 6. Conclusions and discussion

In conclusion, this study has established a comprehensive physics design for the STCF interaction region, covering the selection of key parameters, modular optics design, local chromaticity correction, beam dynamics evaluation, and MDI integration. The resulting design provides a currently optimal solution for the STCF collider rings, forming a core component of the project's conceptual design report.

The selection of critical IR parameters—including the crossing angle $2\theta_c$, the drift length from the IP to the first FF quadrupole (QD1) $L^*$, and beta functions at the IP $\beta_y^*$ and $\beta_x^*$—was guided by a balanced consideration of luminosity performance, chromaticity control, nonlinear dynamics, beam-beam effects, superconducting magnet specifications, and MDI spatial constraints. These parameters fundamentally define both the IR optics and MDI layout.

The modular IR optics design incorporates several features: (1) minimization of the dispersion invariant $H_x$ below 0.02 m throughout the IR to enhance local momentum acceptance (LMA); (2) implementation of exact -I transformations for both CCY and CCX sextupole pairs; (3) specific beta functions at crab sextupoles to reduce crab sextupole strength and nonlinearity; and (4) local chromaticity correction extending to third order to enhance momentum bandwidth. These measures collectively improve resonance driving terms behavior and yield large dynamic and momentum apertures when the IR is integrated into the full ring. Particularly, the minimization of $H_x$ proves highly effective in alleviating the LMA bottleneck, a strategy already implemented in BINP-SCTF [30] and potentially applicable to other colliders such as SuperKEKB, CEPC, and FCC-ee, where enhanced momentum acceptance is essential for achieving adequate beam lifetime.

Fringe fields from the ultra-strong final focusing quadrupoles represent a major source of dynamic aperture reduction in STCF, particularly when interacting with crab sextupoles. Through compensation with strategically placed octupoles near FF quadrupoles, the DA is partially compensated, yielding a Touschek lifetime of 300 s at the luminosity of $9.4\times10^{34}$ cm$^{-2}$s$^{-1}$ and the optimal beam energy of 2 GeV, thereby meeting the STCF design target.

The detector solenoid field is fully compensated using local anti-solenoids between the IP and the first final quadrupole (QD1). This scheme effectively suppresses vertical emittance growth and fully preserves dynamic aperture, confirming that a local perfect compensation strategy should be pursued in practice.

MDI physics evaluations demonstrate compatibility with both IR and detector requirements. Synchrotron radiation from the nearest bending magnet (B0) does not strike the IP beryllium chamber, owing to the optimized beam-pipe transition design and the choice of outer-ring beam incidence (aligning with injection layout). Joint accelerator-detector simulations identify Touschek scattering and beam-gas Coulomb scattering as dominant background sources. Mitigation requires strategic collimator placement to localize particle loss away sensitive regions near the IP, along with an ultra-high dynamic vacuum—approximately $10^{-7}$ Pa in the MDI region and $10^{-8}$ Pa elsewhere—to suppress beam-gas interactions. Technical aspects of the MDI, such as superconducting quadrupole engineering, beam diagnostics, vacuum, collimation, and alignment fall outside the scope of this paper but are detailed in the STCF conceptual design report [1].

To further improve the STCF IR performance, the following optimization priorities are identified: 1) correction of residual second-order vertical and third-order horizontal chromaticity within the IR; 2) optimization of second-order dispersion through tailored dipole length and strength settings in the IR; 3) reduction of chromo-geometric terms arising from breakdown of the -I transformation for off-momentum particles, hopefully via decapole pairs near chromatic sextupole pairs [29]; 4) exploration of fourth- and fifth-order chromaticity correction using octupoles or decapoles at IP image points [29]. Addressing these issues is expected to further improve dynamic and momentum apertures, thereby increasing Touschek lifetime.

In parallel, the following efforts will support the development of the STCF technical design report: (1) Adopting a smaller $\beta_x^*$ (e.g., 40 mm) to help mitigate coherent X-Z instability [51, 52]; (2) Developing flexible IP $\beta$ tuning and robust IP optics correction procedures; (3) Optimizing octupole positions and strengths to further recover DA reduction caused by FF quadrupole fringe fields; (4) Addressing practical challenges in local solenoid compensation, particularly the overlap of residual $B_z$ fields with FF quadrupole fields in the engineering design. Most importantly, the strongly coupled nature of IR optics, FF superconducting quadrupoles, and MDI integration necessitates continued iterative co-optimization in the next design phase.

**Acknowledgements**

We thank A. Bogomyagkov (BINP) for discussions on high-order chromaticity correction and sharing SCTF lattice files. We acknowledge contributions from the MDI team (Wenbin Ma, Xunfeng Li, Mingyi Liu) and beneficial discussions with K. Ohmi (KEK), E. Foresr (KEK) and Z.H. Bai (USTC). This work is supported by the National Natural Science Foundation of China (No. 12341501, No. 12405174). We also thank the Hefei Comprehensive National Science Center for their strong support on the STCF key technology research project.

**References**

[1] PENG Hai-Ping, ZHENG Yang-Heng, ZHOU Xiao-Rong. Super Tau-Charm Facility of China[J]. PHYSICS, 2020, 49(8): 513-524. DOI: 10.7693/wl20200803
[2] P. Raimondi, Status of the SuperB effort, in 2nd Workshop on Super B factory, LNF-INFN, Frascati, 2006, http://www.lnf.infn.it/conference/superb06/talks/raimondi1.ppt.
[3] M. Zobov, et al. Test of crab-waist collisions at DAΦNE Φ factory. Phys. Rev. Lett. 104, 174801 (2010).
[4] P. Raimondi, M. Zobov, and D. Shatilov. Suppression of beam-beam resonances in crab waist

collisions. In Proceedings of EPAC08, Genoa, Italy, 2008.

[5] CEPC Study Group Collaboration, CEPC Technical Design Report: Accelerator. Radiat. Detect. Technol. Methods 8, 1 (2024) pp.1-1105.

[6] A. Abada, M. Abbrescia, S.S. AbdusSalam, et al. FCC-ee: the Lepton Collider: Future Circular Collider Conceptual Design Report. Eur. Phys. J. Spec. Top. 228(2), 261–623 (2019).

[7] M. E. Biagini, R. Boni, M. Boscolo, et al. SuperB Progress Reports -- The Collider. arXiv:1008.1541. http://arxiv.org/abs/1009.6178v3 https://doi.org/10.48550/arXiv.1009.6178

[8] Y. Ohnishi, Recent performance of SuperKEKB. at the 70th ICFA Advanced Beam Dynamics Workshop on High Luminosity Circular e+e- Colliders (eeFACT 2025), Tsukuba, Japan, March 3-7 2025.

[9] Epifanov, D.A., SCTF Collaboration. Project of Super Charm-Tau Factory. Phys. Atom. Nuclei 83, 944–948 (2020). https://doi.org/10.1134/S1063778820060137

[10] Jingyu Tang, Project status and R&D efforts for Super Tau-Charm Facility, at the 70th ICFA Advanced Beam Dynamics Workshop on High Luminosity Circular e+e- Colliders (eeFACT 2025), Tsukuba, Japan, March 3-7 2025.

[11] M.E. Biagini, R. Boni, M. Boscolo, et al. Tau/Charm Factory Accelerator Report. INFN Report INFN-13-13/LNF, September 2013. https://doi.org/10.48550/arXiv.1310.6944

[12] A. Bogomyagkov, E. Levichev, and P. Piminov. Final focus designs for crab waist colliders. Phys. Rev. Accel. Beams, 19, 121005 (2016).

[13] K. Ohmi, and H. Koiso. Dynamic aperture limit caused by IR nonlinearities in extremely low-beta B factories. Proceedings of IPAC'10, Kyoto, Japan, 2010.

[14] Y. Ohnishi, T. Abe, T. Adachi, et al. Accelerator design at SuperKEKB. Prog. Theor. Exp. Phys. 2013, 03A011 (17 pages). DOI: 10.1093/ptep/pts083

[15] Y. Ohnishi, T. Abe, K. Akai, et al. SuperKEKB operation using crab waist collision scheme. Eur. Phys. J. Plus (2021) 136:1023. https://doi.org/10.1140/epjp/s13360-021-01979-8

[16] W. W. Gao et al. Interaction section lattice design for a STCF project. In Proceedings of the 10th International Particle Accelerator Conference 269 (IPAC-2019), Melbourne, Australia, 2019.

[17] J. Q. Lan et al. Design of beam optics for a Super Tau-Charm Factory. Journal of Instrumentation, 16: T07001, 2021.

[18] Tao Liu, Chun Zhang, Lin Wang et al. Design of a hybrid seven-bend-achromat-based lattice for a Super Tau Charm Facility. In proceedings of 14th International Particle Accelerator Conference (IPAC2023), Venice, Italy, 2023.

[19] T. Liu, L.H. Zhang, S.Y. Li, et al. Recent progress and future plan for STCF collider ring lattice design. Modern Physics Letters A, 39:2440005, 2024.

[20] Y. Zou, L.H. Zhang, T. Liu, et al. Optics design of the Super Tau-Charm Facility collider rings. Nuclear Instruments and Methods in Physics Research Section A, 2025 (To be published). https://doi.org/10.48550/arXiv.2507.18924

[21] Xiao-Cong Ai, et al., Conceptual Design Report of Super Tau-Charm Facility: The Accelerator, (2025), Nucl. Sci. Tech. 36, 242 (2025)/arXiv: 2509.11522 [physics.acc-ph]

[22] K. Ohmi, N. Kuroo, K. Oide, et al. Coherent Beam-Beam Instability in Collisions with a Large Crossing Angle. Phys. Rev. Lett. 119, 134801 (2017).

[23] L.H. Zhang, T. Liu, S.Y. Li, et al., Longitudinal beam dynamics and collective effects at the STCF collider rings. Modern Physics Letters A, 39:24440008, 2024.

[24] Y. Ohnishi, Recent progress of SuperkekB project and future prospect. In proceedings of 14th

International Particle Accelerator Conference (IPAC2023),Venice, ltaly, 2023.

[25] V. I. Telnov, Restriction on the Energy and Luminosity of e+e- Storage Rings due to Beamstrahlung, Phys. Rev. Lett. 110, 114801 (2013).

[26] T. O. Raubenheimer, and F. Zimmermann, Final-focus systems in linear colliders. Rev. Mod. Phys. 72, 95 (2000).

[27] H. Koiso, A. Morita, Y. Ohnishi, et al. Lattice of the KEKB colliding rings. Prog. Theor. Exp. Phys. 2013, 03A009. DOI: 10.1093/ptep/pts086

[28] PEP-II: An Asymmetric B factory. Conceptual Design Report. Report No.: LBL-PUB-5379, SLAC-418, CALT-68-1869, UCRL-ID-114055, UC-IIRPA-93-01, June, 1993.

[29] P. Raimondi, Simone Maria Liuzzo, and Michael Hofer. Local chromatic correction optics for Future Circular Collider e+e-. Phys. Rev. Accel. Beams, 28:021002, 2025.

[30] A. Bogomyagkov, Main challenges of the Crab Waist Super Charm Tau Factory, in the 6th International Workshop on Future Tau Charm Facilitis, Guangzhou, China, November 17-21, 2024.

[31] A. Bogomyagkov, Touschek lifetime optimization at Super C-Tau factory in Novosibirsk, The 2024 International Workshop on Future Tau Charm Facilities (FTCF2024), Hefei, China, 14–18 January, 2024.

[32] K. Oide, M. Aiba, S. Aumon, et al. Design of beam optics for the future circular collider e+e- collider rings. Phys. Rev. Accel. Beams, 19:111005, 2016.

[33] C Steier, D Robin, L Nadolski, et al. Measuring and optimizing the momentum aperture in a particle accelerator, Physical Review E, vol. 65, 056506, 2002. doi: 10.1103/PhysRevE.65.056506.

[34] K. Brown, R. Servranckx, First- and second-order charged particle optics, SLAC-PUB-3381, 1984.

[35] M. Donald, R. Helm, J. Irwin, H. Moshammer, et al., Localized Chromaticity Correction of Low-Beta Insertions in Storage Rings, SLAC-PUB-6197, April 1997.

[36] G. Xu, General conditions for self-cancellation of geometric aberrations in a lattice structure. Phys. Rev. ST Accel. Beams 8, 104002 (2005).

[37] Yunhai Cai, Analytical approach to chromatic correction in the final focus system of circular colliders. Phys. Rev. ST Accel. Beams 19, 111006 (2016).

[38] E. Forest, D. Robin, A. Zholents, et al. Local Chromatic Correction Scheme for LER of PEP-II. In the International Workshopon B-Factories: Accelerators and Experiments, KEK, Tsukuba, Japan, Nov. 17-20, 1992. Report number: SLAC-REPRINT-1993-042, LBL-33479.

[39] R. Brinkmann, Optimization of a Final Focus System for Large Momentum Bandwidth. DESY Report No. M-90-14, November 1990.

[40] B. W. Montague, Linear optics for improved chromaticity correction, Report No. CERN-LEP-NOTE-165, 1979.

[41] Lingyun Yang, Yongjun Li, Weiming Guo, and Samuel Krinsky, Multiobjective optimization of dynamic aperture, Phys.Rev.STAccel.Beams14,054001(2011).

[42] A. Franchi, R. Tomás, and F. Schmidt, Magnet strength measurement in circular accelerators from beam position monitor data, Phys. Rev. ST Accel. Beams 10, 074001 (2007).

[43] T. Liu. Pamkit. https://pypi.org/project/PAMKIT/. Accessed: 2025-07-22.

[44] SAD home page. https://acc-physics.kek.jp/SAD/

[45] L.H. Zhang, Optics design of the interaction region for the STCF, at the 70th ICFA Advanced

Beam Dynamics Workshop on High Luminosity Circular e+e- Colliders (eeFACT 2025), Tsukuba, Japan, March 3-7 2025.

[46] A. V. Bogomyagkov, Eu. B. Levichev, P. A. Piminov, et al., "Analysis of the Non-Linear Fringe Effects of Large Aperture Triplets for the HL-LHC Project," in Proceedings of IPAC2013, Shanghai, China.

[47] Ande Ma, Yiwei Wang, Yuan Zhang, et al. Minimization of the vertical emittance growth caused by the solenoid field in the interaction region of a circular collider. JINST 20 P04034 (2025).

[48] M. Sullivan, S. Ecklund, J. Seeman, et al., The Beam-Stay-Clear Definition of the PEP-II B Factory. Report NO.: SLAC-PUB-7207, June 1996. https://www.slac.stanford.edu/pubs/slacpubs/7000/slac-pub-7207.pdf

[49] H.C. Shi, B. Wang, H.Y. Shi, et al. Beam background simulation and experiment at BEPCII. Nuclear Inst. and Methods in Physics Research, A 1050 (2023) 168174.

[50] A. Natochii, S. E. Vahsen, H. Nakayama, et al. Improved simulation of beam backgrounds and collimation at SuperKEKB. PHYS. REV. ACCEL. BEAMS 24, 081001 (2021).

[51] Sangya Li, Demin Zhou, Linhao Zhang, et al. Investigating luminosity optimization in STCF with crab waist scheme by beam-beam simulation. Modern Physics Letters A Vol. 39, No. 40 (2024) 2440013.

[52] Peter Kicsiny, Demin Zhou, Xavier Buffat, et al., Incoherent horizontal emittance growth due to the interplay of beam-beam interaction and longitudinal wakefield in crab-waist colliders. PHYS. REV. ACCEL. BEAMS 28, 051002 (2025).

**Appendix: Optics design at the crab sextupole**

Following K. Oide's derivation for the strength of the crab sextupole (CS) [31], we present a more detailed derivation and analysis to guide the optics design at the CS in our scenario.

As shown in Fig. 17, taking the positron beam as an example, the crab-waist scheme relates the twisting distance $\Delta s$ of the vertical $\beta_y$ beam waist at the IP to the horizontal displacement $x^*$ via:

$$\Delta s = -\frac{x^*}{\tan(2\theta_c)}, \tag{A1}$$

where the negative sign indicates that crab-waist twists the vertical $\beta_y$ beam waist of particles with positive $x^*$ in the direction opposite to the beam's motion. Therefore, the relationship between the particle coordinates after twisting and the particle coordinates at the IP is:

$$\begin{pmatrix} y \\ p_y \end{pmatrix} = \begin{bmatrix} 1 & \Delta s \\ 0 & 1 \end{bmatrix} \begin{pmatrix} y^* \\ p_y^* \end{pmatrix}, \tag{A2}$$

and the associated transformation is:

$$y = y^* + p_y^* \Delta s = y^* - \frac{x^* p_y^*}{\tan(2\theta_c)}, \tag{A3}$$

which means that the Crab-Waist at the IP causes $y^*$ to change by:

$$\Delta y = -\frac{x^* p_y^*}{\tan(2\theta_c)}. \tag{A4}$$

To achieve this $\Delta y$, the required Hamiltonian is:

$$\frac{\partial H_{CW}}{\partial p_y^*} = \Delta y = -\frac{x^* p_y^*}{\tan(2\theta_c)} \Rightarrow H_{CW} = -\frac{x^* p_y^{*2}}{2\tan(2\theta_c)}. \tag{A5}$$

This Hamiltonian required by the crab-waist scheme can be provided by a sextupole (the so-called crab sextupole). The Hamiltonian of a normal sextupole is:

$$H_s = \frac{K_2}{6}\left(x^3 - 3xy^2\right), \tag{A6}$$

with $K_2$ the integral strength of the sextupole. Using the well-known linear transfer matrix from $s_1$ to $s_2$ expressed by the twiss parameters and considering $\alpha^*_{x/y} = 0$ at the IP, the coordinates $x/y$ at the crab sextupole can be expressed by the coordinates at the IP as:

$$x = x^* \sqrt{\frac{\beta_{x,CS}}{\beta_x^*}} \cos\Delta\mu_x + p_x^* \sqrt{\beta_{x,CS}\beta_x^*} \sin\Delta\mu_x, \tag{A7}$$

$$y = y^* \sqrt{\frac{\beta_{y,CS}}{\beta_y^*}} \cos\Delta\mu_y + p_y^* \sqrt{\beta_{y,CS}\beta_y^*} \sin\Delta\mu_y, \tag{A8}$$

Substituting Eqs. (A7) and (A8) into the sextupole Hamiltonian Eq. (A6) gives:

$$H_s = -\frac{K_2}{2}\left(x^* \sqrt{\frac{\beta_{x,CS}}{\beta_x^*}} \cos\Delta\mu_x + p_x^* \sqrt{\beta_{x,CS}\beta_x^*} \sin\Delta\mu_x\right) \cdot$$
$$\left(y^* \sqrt{\frac{\beta_{y,CS}}{\beta_y^*}} \cos\Delta\mu_y + p_y^* \sqrt{\beta_{y,CS}\beta_y^*} \sin\Delta\mu_y\right)^2$$
$$+ \frac{K_2}{6}\left(x^* \sqrt{\frac{\beta_{x,CS}}{\beta_x^*}} \cos\Delta\mu_x + p_x^* \sqrt{\beta_{x,CS}\beta_x^*} \sin\Delta\mu_x\right)^3 \tag{A9}$$

By appropriately setting the phase advance from the crab sextupole to the IP, i.e., $\Delta\mu_x = m\pi$, and $\Delta\mu_y = n\pi + \pi/2$ (to eliminate undesired nonlinear terms), the Hamiltonian simplifies to:

$$H_S = -\frac{K_2}{2}\text{sgn}(\cos\Delta\mu_x)\sqrt{\frac{\beta_{x,CS}}{\beta_x^*}}\beta_{y,CS}\beta_y^* x^* p_y^{*2} + \frac{K_2}{6}\left(x^* \sqrt{\frac{\beta_{x,CS}}{\beta_x^*}}\text{sgn}(\cos\Delta\mu_x)\right)^3, \tag{A10}$$

Comparing Eq.(A10) with the required Crab-Waist Hamiltonian Eq.(A5), the integrated strength $K_2$ of the crab sextupole is derived as:

$$K_2 = \frac{1}{\tan(2\theta_c)\beta_{y,CS}\beta_y^*}\sqrt{\frac{\beta_x^*}{\beta_{x,CS}}}\frac{1}{\text{sgn}(\cos\Delta\mu_x)}, \tag{A11}$$

Clearly, once the key IR parameters are fixed, increasing $\beta_x$ and $\beta_y$ functions at the crab sextupole can reduce its strength $K_2$, with a larger $\beta_y$ being more effective than a larger $\beta_x$ for lowering $K_2$. On the other hand, to minimize the adverse effects from the crab sextupole (e.g. the second term in Eq.(A10)), $\beta_x$ function at the crab sextupole should be designed as small as possible. However, an excessively small $\beta_x$ would undesirably increase the horizontal chromaticity. Therefore, the optimized beta functions at the crab sextupole should satisfy $\beta_y \gg \beta_x$ while maintaining $\beta_x$ as small as possible but larger than 1.

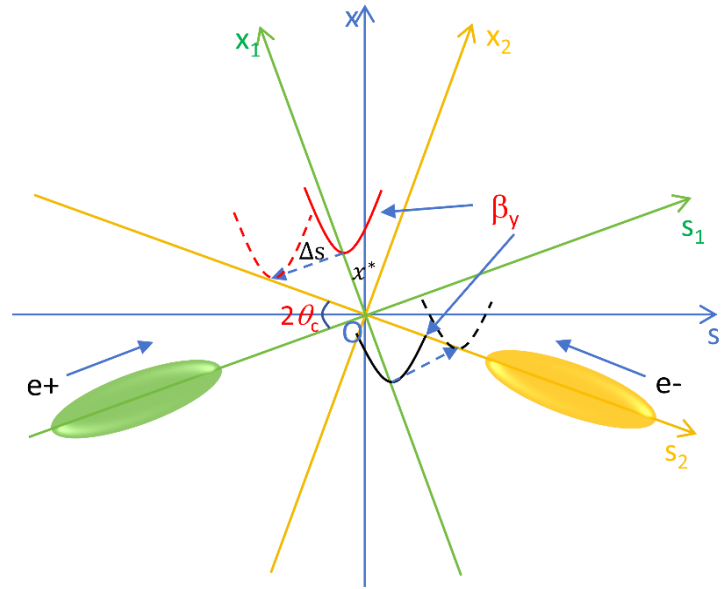

Figure 17: The crab-waist scheme shifts the vertical waist of a beam in s-direction by Δs onto the center of the other beam.